\documentclass[%
reprint,
superscriptaddress,
amsmath,amssymb,
aps,
prl
]{revtex4-1}

\usepackage{tabu}
\usepackage{graphicx}
\usepackage{epstopdf}
\usepackage{dcolumn}
\usepackage{bm}
\usepackage[breaklinks=true]{hyperref}
\hypersetup{colorlinks=true, citecolor=blue, filecolor=blue, linkcolor=blue,urlcolor=black}
\usepackage{hyperref}

\begin{document}

\preprint{APS/123-QED}

\title{Microwave control of thermal magnon spin transport}

\author{J. Liu}
\thanks{}

\email{jing.liu@rug.nl}
\affiliation{Physics of Nanodevices, Zernike Institute for Advanced Materials, University of Groningen, Nijenborgh 4, 9747 AG Groningen, Netherlands}
\author{F. Feringa}
\affiliation{Physics of Nanodevices, Zernike Institute for Advanced Materials, University of Groningen, Nijenborgh 4, 9747 AG Groningen, Netherlands}%
\author{B. Flebus}
\affiliation{Department of Physics and Astronomy, University of California, Los Angeles, California 90095, USA}%
\author{L.J. Cornelissen}
\affiliation{Physics of Nanodevices, Zernike Institute for Advanced Materials, University of Groningen, Nijenborgh 4, 9747 AG Groningen, Netherlands}%
\author{J.C. Leutenantsmeyer}
\affiliation{Physics of Nanodevices, Zernike Institute for Advanced Materials, University of Groningen, Nijenborgh 4, 9747 AG Groningen, Netherlands}%
\author{R.A. Duine}
\affiliation{Institute for Theoretical Physics, University of Utrecht, Princetonplein 5, 3584 CC Utrecht, The Netherlands}%
\affiliation{Department of Applied Physics, Einhoven University of Technology, PO Box 513, 5600 MB Eindhoven, The Netherlands}
\author{B.J. van Wees}
\affiliation{Physics of Nanodevices, Zernike Institute for Advanced Materials, University of Groningen, Nijenborgh 4, 9747 AG Groningen, Netherlands}%

\begin{abstract}
	
\textbf{We observe that an rf microwave field strongly influences the transport of incoherent thermal magnons in yttrium iron garnet. Ferromagnetic resonance in the nonlinear regime suppresses thermal magnon transport by 95$\%$. The transport is also modulated at non-resonant conditions in two cases, both related to the magnon band minimum. Firstly, a strong enhancement of the nonlocal signal appears at a static magnetic field below the resonance condition. This increase only occurs at one field polarity and can be as large as 800$\%$. We attribute this effect to magnon kinetic processes, which give rise to band-minimum magnons and high-energy chiral surface modes. Secondly, the signal increases at a static field above the resonance condition, where the rf frequency coincides with the magnon band minimum. Our study gives insight into the interplay between coherent and incoherent spin dynamics: The rf field modifies the occupation of relevant magnon states and, via kinetic processes, the magnon spin transport.}

\end{abstract}

\maketitle

\begin{figure}[tb]
	\centering
	\includegraphics[width=0.9\linewidth]{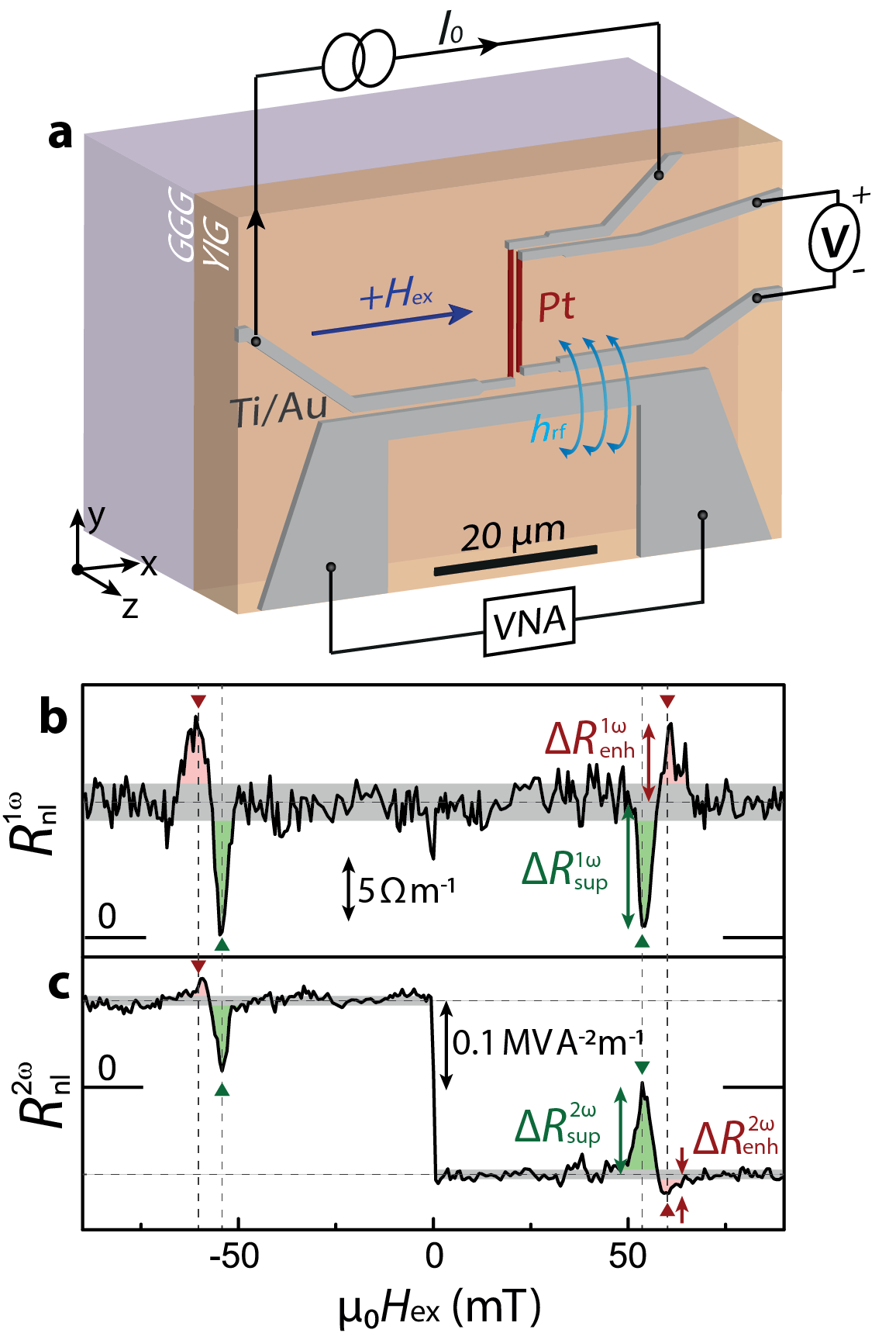}
	\caption{\textbf{Experiment schematic and typical results.} \textbf{a} The YIG film lies in the xy-plane on top of GGG substrate. Pt (red) strips are along the y-axis. Ti/Au leads (grey) contacted to Pt strips are connected with measurement setups. An ac current with rms value of $I_{0}$ is sent through the left Pt strip. Using a lockin-technique, we measured the first and second harmonic voltages by the right Pt strip. The Ti/Au structure on the bottom side is a shorted end of a coplanar stripline where an rf current is driven through by a VNA, resulting in an rf magnetic field $\textit{\textbf{h}}_{\textrm{rf}}$. An external magnetic field $\textit{H}_{\textrm{ex}}$ is applied along the x-axis with positive sign corresponding to the positive x direction. Field-dependent \textbf{b} first and \textbf{c} second harmonic nonlocal resistances. At resonance (green triangles), both first and second harmonic signals are suppressed, indicated by the green areas with magnitudes defined as $\Delta R^{1\omega}_{\textrm{sup}}$ and $\Delta R^{2\omega}_{\textrm{sup}}$. At a higher field (red triangles), there is an enhancement of the signals (red areas) with magnitudes of $\Delta R^{1\omega}_{\textrm{enh}}$ and $\Delta R^{2\omega}_{\textrm{enh}}$. The grey bar represents the baseline resistance with a width corresponding to its standard deviation. The injector-to-detector distance is $1\,\mu$m on top of a 210$\,$nm thick YIG. A continuous rf power is applied: $P_{\textrm{rf}}=+19\,$dBm, $\omega_{\textrm{rf}}/2\pi=3\,$GHz.} 
	\label{fig1}
\end{figure}

Thermal magnons are intrinsic fluctuations of the magnetization in a magnet. Non-equilibrium thermal magnons with a small deviation from equilibrium can be described by a temperature and a chemical potential \cite{PhysRevB.94.014412}. They can be generated by a temperature gradient, a process known as the spin Seebeck effect (SSE) \cite{uchida2010observation}. Moreover, they can be electrically excited \cite{zhang2012spin,zhang2012magnon,cornelissen2015long,goennenwein2015non,li2016observation,wu2016observation,velez2016competing, shan2017nonlocal, ganzhorn2017non} and diffusively propagate under a gradient of the magnon chemical potential, $\nabla\mu_{\textrm{m}}$, with a diffusion length as long as 10$\,\mu$m at room temperature \cite{cornelissen2015long}. These magnons have energy up to $k_{\textrm{B}}T\sim6\,$THz, where exchange energy dominates. Lately, a lot of effort has been made to control the transport of these electrically excited incoherent high-energy magnons \cite{liu2017magnon,cornelissen2018spin, ganzhorn2016magnon}, because they open up a new way of miniaturizing magnonic devices, due to their short wavelength and their dc-current controllable character. Recently, long-distance electrically-controlled propagation has also been realized in an antiferromagnet \cite{lebrun2018tunable}.

In contrast, coherent magnons have well-defined frequency and long wavelength. They can propagate over long distances on the order of centimeters as coherent waves, which is appealing for logic implementation in magnonic devices \cite{kruglyak2010magnonics,chumak2015magnon,chumak2014magnon}. They can be excited by a microwave field, in a fashion depending on the relative orientations of the rf field ($\textit{\textbf{h}}_{\textrm{rf}}$) and magnetization ($\textit{\textbf{M}}$): When $\textit{\textbf{h}}_{\textrm{rf}}\perp\textit{\textbf{M}}$, a uniform precession mode, known as ferromagnetic resonance (FMR) \cite{kittel1948theory,chumak2015magnon,kruglyak2010magnonics}, can be excited. For $\textit{\textbf{h}}_{\textrm{rf}}\parallel\textit{\textbf{M}}$, parametric pumping \cite{sandweg2011spin} can be realized. Generally, for a ferro- or ferrimagnetic insulator, an rf field driving it into FMR oscillates at GHz frequency, where magnetic dipole interactions dominate. Alternatively, coherent magnons with THz frequency can be excited by femtosecond laser pulses \cite{seifert2018femtosecond,bocklage2017coherent}. Moreover, spin orbit torque (SOT) can also generate coherently propagating magnons \cite{collet2016generation}.

The dispersive properties of magnons in thermal equilibrium can be described by the dipole-exchange spin wave spectrum \cite{kalinikos1990dipole,barker2016thermal}. Due to their bosonic nature, the distribution obeys Bose-Einstein statistics. The interplay between incoherent thermal magnons and coherent magnons has been under debate for a decade \cite{flebus2016local,l1990nonlinear,kruglyak2010magnonics,thiery2018nonlinear,du2017control} and its better understanding would also lead to crucial insights into magnon Bose-Einstein condensation (BEC), which has already been observed at room temperature \cite{demokritov2006bose}.

Here, we study the transport of electrically injected and Joule-heating induced thermal magnons in the presence of an rf field. We find that for an in-plane magnetization, the rf power can have a strong influence on the transport in a few special situations: (i) At the onset of kinetic processes, which give rise to a large population of band-minimum magnons and higher-energy magnons with chiral surface mode character; (ii) at the ferromagnetic resonance (FMR) condition; (iii) when the rf field oscillates at the frequency of the band-minimum magnons.  \noindent

\begin{flushleft}
	\textbf{RESULTS}
\end{flushleft}
{\textbf{Experiment.} The devices are fabricated on 210$\,$nm and 100$\,$nm thick single crystal yttrium iron garnet (YIG) films. Fig.$\,$\ref{fig1}a shows a schematic of a typical device: two 7$\,$nm thick Pt strips on YIG are contacted to Ti$|$Au leads for electrical connection. An on-chip stripline with a shorted end is also made of the Ti$|$Au layer. The stripline is connected to a vector network analyser (VNA), which sends a high-frequency ac current through the line and generates the rf field $\textit{\textbf{h}}_{\textrm{rf}}$, mostly out of the film plane at the Pt device. An external static field $\textbf{\emph{H}}_{\textrm{ex}}$ is applied to align the magnetization of YIG in the film plane perpendicular to the Pt strips. Devices on 210$\,$nm and 100$\,$nm thick YIG have injector-to-detector distances of 1$\,\mu$m and 600$\,$nm, respectively. Details about the sample preparation and experimental setup are given in METHODS.

To study the transport of thermal magnons, we conduct a nonlocal measurement, where we send an ac current through one of the Pt strips (injector) as shown in Fig.$\,$\ref{fig1}a. This ac current oscillates at a frequency less than 20$\,$Hz, which is quasi-dc comparing with the frequencies of the current through the stripline, 3$\,$GHz, 6$\,$GHz and 9$\,$GHz. Via the spin Hall effect (SHE) in Pt, electron spin accumulation at the Pt$|$YIG interface leads to nonequlibrium magnon spin accumulation. Under a gradient of the magnon chemical potential, magnon spins diffuse towards the other Pt strip (detector), where the reciprocal process takes place, namely magnon electrical detection. Magnon spin current converts back into an electron spin current, which can be measured as an electrical voltage via the inverse spin Hall effect (ISHE). Meanwhile, the Joule heating associated with the current passing through the Pt injector, $\sim I_{0}^{2}R$, induces a thermal gradient in YIG, which drives a magnon spin current both horizontally and vertically \cite{shan2016influence}. This causes a nonequilibrium magnon accumulation and depletion near the Pt injector and at the bottom of the YIG film, respectively, which both propagate under a gradient of magnon chemical potential. Depending on the ratio of injector-to-detector distance to the YIG thickness, one will dominate. We refer to this process as thermal magnon injection. The resulting nonequilibrium magnon spin current can also be detected by the Pt detector via ISHE. Due to the competition between the magnon spin currents with opposite sign coming from the top and bottom of YIG in the vicinity of the injector, the measured ISHE voltage changes sign as a function of the injector-to-detector distance \cite{shan2016influence}. The results shown in this paper are all in the regime of large injector-to-detector distance. 

With a lockin technique, we separately study the magnon spin transport resulting from the electrical and thermal injection by measuring the first and second harmonic nonlocal voltages ($V^{1\omega}_{\textrm{nl}}$ and $V^{2\omega}_{\textrm{nl}}$), which are recorded at the same ($\omega_{\textrm{lockin}}$) and double frequency ($2\omega_{\textrm{lockin}}$) of the excitation current $I_{0}$. We define the first and second harmonic nonlocal "resistance" as 
\begin{align}
	R^{1\omega}_{\textrm{nl}}&=V^{1\omega}_{\textrm{nl}}/(I_{0}L),\\
	R^{2\omega}_{\textrm{nl}}&=V^{2\omega}_{\textrm{nl}}/(I_{0}^{2}L),
\end{align}
where we not only normalize the nonlocal voltage by the current but also by the the length of the device $L$, since the ISHE voltage scales with it. We study $R^{1\omega}_{\textrm{nl}}$ and $R^{2\omega}_{\textrm{nl}}$ as a function of the static field $H_{\textrm{ex}}$ in the presence of a continuous rf power, which generates an rf field. We align $H_{\textrm{ex}}$ perpendicular to the Pt strip by eye to achieve the highest magnon injection and detection efficiency. Typical rms-amplitude and frequency of the current are: $I_{0}=200\,\mu$A, $\omega_{\textrm{lockin}}/2\pi=17.777\,$Hz. Three different rf frequencies ($\omega_{\textrm{rf}}/2\pi$) are used: 3$\,$GHz, 6$\,$GHz and 9$\,$GHz, and applied rf powers ($P_{\textrm{rf}}$) range from $-10\,$dBm to $+19\,$dBm (0$\,$dBm$=$1$\,$mW). Moreover, we read the reflected rf power ($S_{11}$ parameter) from VNA to monitor the global magnetization dynamics. The experiments are conducted at room temperature in atmosphere.

\noindent\textbf{Nonlocal signals under an rf field.} In Fig.$\,$\ref{fig1}b, typical field-dependent results show that the first harmonic nonlocal resistance ($R^{1\omega}_{\textrm{nl}}$) has a baseline of $\sim8\,\Omega\,\textrm{m}^{-1}$  as a result of the magnon injection and detection, while the second harmonic signal ($R^{2\omega}_{\textrm{nl}}$) in Fig.$\,$\ref{fig1}c reverses its sign by changing the polarity of the magnetization due to the opposite sign of the magnon spins. We define the magnitude of the second harmonic signal as half of the difference between the baseline nonlocal second harmonic resistances at positive and negative static field, which is $\sim0.1 \,\textrm{MV}\,\textrm{A}^{-2}\textrm{m}^{-1}$. At FMR we observe a suppression in both first and second harmonic signals, defined as $\Delta R^{1\omega}_{\textrm{sup}}$ and $\Delta R^{2\omega}_{\textrm{sup}}$. Besides, at a higher field close to the resonance condition there is an enhancement of the signals, denoted as $\Delta R^{1\omega}_{\textrm{enh}}$ and $\Delta R^{2\omega}_{\textrm{enh}}$. To exclude any possible thermal effects that might influence the nonlocal signals at high pumping power, we perform local measurements, where we send the current through one Pt strip and measure the voltage across the same strip (see METHODS and Supplementary Fig.$\,$S4). Comparing the local and nonlocal measurements we find that the influence of the rf field on voltages created at the individual injector and detector Pt strips is marginal compared with the influence of the rf field on voltages due to the magnon spin transport. Besides, we confirm that the first harmonic signals scale linearly with the excitation current (see Supplementary Fig.$\,$S6) and are not changed by using different lockin-frequencies for both low and high rf powers. This ensures that the measured nonlocal resistance is due to the magnon spin transport from the electrical magnon injection instead of other spurious effects caused by the microwave power \cite{iguchi2016measurement}.

\begin{figure}[tb]
	\centering
	\includegraphics[width=0.65\linewidth]{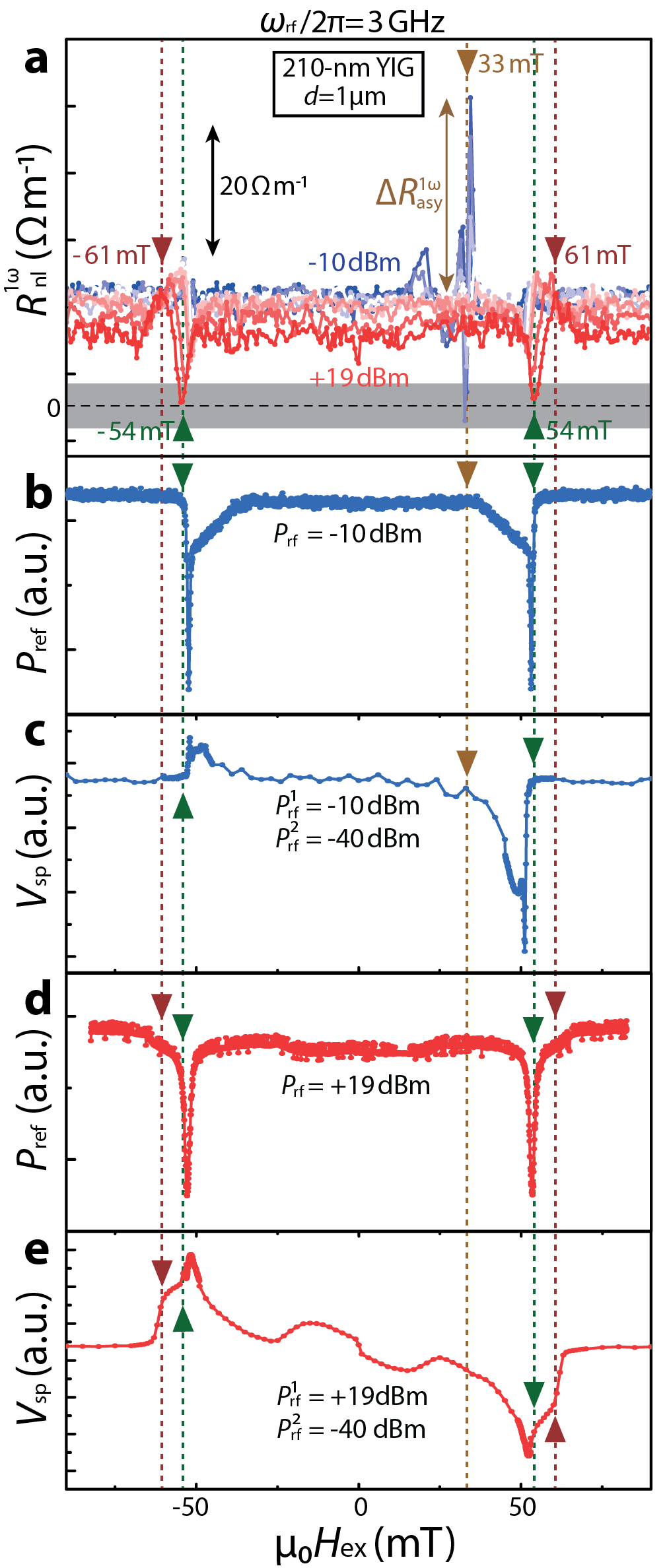}
	\caption{\textbf{Comparing nonlocal results with microwave power reflection and spin-pumping measurement.} \textbf{a} Field dependent first harmonic nonlocal signals at different rf powers. The highest and lowest applied rf powers are -10$\,$dBm (blue) and 19$\,$dBm (red) in \textbf{a}. The scale bar is 20$\,\Omega\,$m$^{-1}$. The brown, green and red triangles with corresponding vertical dashed lines indicate special field positions. Field-dependent reflected microwave powers at \textbf{b} low and \textbf{d} high rf power, and spin-pumping ISHE voltage at \textbf{c} low and \textbf{e} high rf power. 
	}
	\label{dispersion_for_three_fields_two_thicknesses}
\end{figure}

\begin{figure*}[tb]
	\centering
	\includegraphics[width=0.99\linewidth]{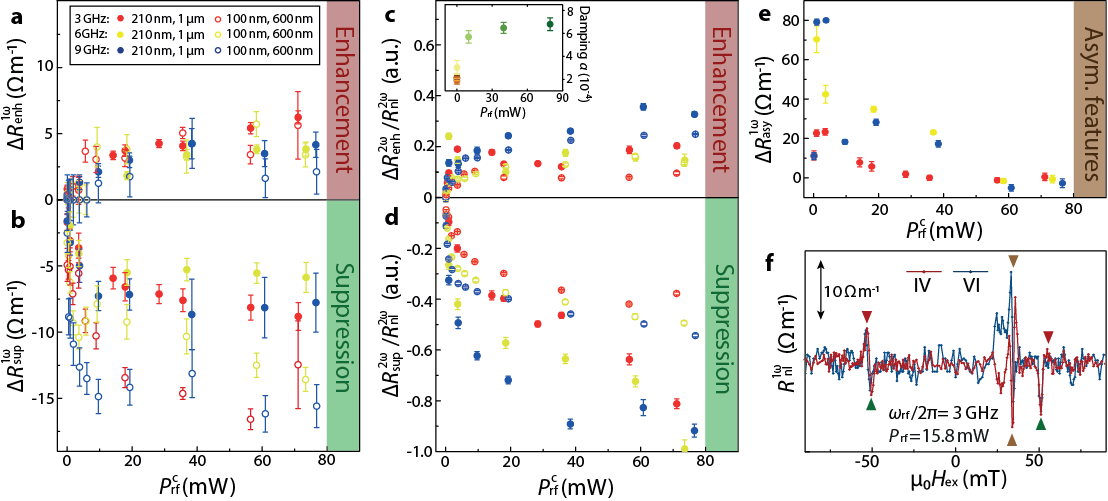}
	\caption{\textbf{Suppression and enhancement of nonlocal signals as a function of calibrated rf power.} Rf-power dependent \textbf{a} $R_{\textrm{enh}}^{1\omega}$, \textbf{b} $R_{\textrm{sup}}^{1\omega}$, \textbf{c} $R_{\textrm{enh}}^{2\omega}/R_{\textrm{nl}}^{2\omega}$, \textbf{d} $R_{\textrm{sup}}^{2\omega}/R_{\textrm{nl}}^{2\omega}$ and \textbf{e} $R_{\textrm{asy}}^{1\omega}$ at different rf frequencies (3/6/9$\,$GHz: red/yellow/blue) with different YIG thicknesses (210$\,$nm/100$\,$nm: solid/open dots) and corresponding injector-to-detector distance of 1$\,\mu$m and 600$\,$nm, respectively. Same color and style code is used for the rf frequency and YIG thickness in \textbf{a}-\textbf{e}. The inset of \textbf{c} is the effective Gilbert damping parameter $\alpha$ as a function of $\textit{P}_{\textrm{rf}}$. Some error bars are smaller than the symbols. \textbf{f} The reciprocity check has been performed by interchanging the injector and detector. Two measurement schemes are noted as IV- and VI-configuration. }
	\label{fig4}
\end{figure*}

\noindent\textbf{Rf-power dependency.} The influence of the rf field on the nonlocal transport strongly depends on the input power to generate the rf field ($P_{\textrm{rf}}$). In Fig.$\,$\ref{dispersion_for_three_fields_two_thicknesses}a and $\,$\ref{isofrequencycurves}a, we show typical nonlocal results at different rf powers for rf frequencies of 3$\,$GHz and 6$\,$GHz, respectively. Surprisingly, when the rf power is relatively low, the nonlocal signal changes significantly at a static field lower than the FMR condition. This feature only appears at a positive static field. This should be contrasted with the situation at high powers (above $\sim10\,$mW), where the suppression and enhancement of the nonlocal signals at the FMR condition and at a higher field than FMR arise for both positive and negative field: the example in Fig.$\,$\ref{fig1}b shows the highest-power case. Besides, in the second harmonic signals we see these asymmetric features, albeit less prominently, which is probably due to the different magnon injection mechanism: In the thermally injected case for the second harmonic signals, the temperature gradient in a thin film complicates the physical scenario \cite{shan2016influence}. Therefore, we focus on the results of the first harmonic signals. We record the magnitude of the asymmetric enhancement at positive static fields as $\Delta R^{1\omega}_{\textrm{asy}}$ as shown in Fig.$\,$\ref{dispersion_for_three_fields_two_thicknesses}a and $\,$\ref{isofrequencycurves}a, which is plotted against rf power in Fig.$\,$\ref{fig4}e. In the case of 6$\,$GHz rf frequency shown in Fig.$\,$\ref{isofrequencycurves}a, $\Delta R_{\textrm{asy}}^{1\omega}$ can be more than 8 times larger than the baseline resistance of $\textit{R}_{\textrm{nl}}^{1\omega}$. This asymmetric feature has also been observed with different rf frequencies on YIG with different thicknesses. $\Delta R_{\textrm{asy}}^{1\omega}$ increases drastically with rf power until $P_{\textrm{rf}}\sim10\,$mW, where it starts to decrease. Moreover, when we interchange the role of injector and detector, the asymmetric feature changes as shown in Fig.$\,$\ref{fig4}f, whereas the features at FMR ($\Delta R_{\textrm{sup}}^{1\omega}$) and at a higher field than FMR ($\Delta R_{\textrm{enh}}^{1\omega}$) remains almost the same. Besides, comparing results of 210$\,$nm and 100$\,$nm thick YIG, this asymmetric feature is much more significant in the 210$\,$nm thick YIG (see Supplementary Fig.$\,$S8).

We summarize the amplitudes of the suppression and enhancement of the nonlocal signals, $\Delta R_{\textrm{sup}}^{1\omega}$ ($\Delta R_{\textrm{sup}}^{2\omega}$) and $\Delta R_{\textrm{enh}}^{1\omega}$ ($\Delta R_{\textrm{enh}}^{2\omega}$), as a function of the delivered rf power $P_{\textrm{rf}}^{\textrm{c}}$ for different rf frequencies and both YIG thicknesses in Fig.$\,$\ref{fig4}a-\ref{fig4}d, noting a drastic increase with rf power, until at $\sim15\,$mW there is no significant increase anymore. A similar trend can be seen from the damping parameter of the FMR mode as shown in the inset of Fig.$\,$\ref{fig4}c. Other rf power dependent results (210$\,$nm and 100$\,$nm YIG; 3$\,$GHz and 9$\,$GHz rf frequencies) can be found in Supplementary Fig.$\,$S8 and the data extraction method is provided in METHODS and Supplementary Fig.$\,$S3. The rf power calibration is provided in METHODS. The damping parameters in the inset of Fig.$\,$\ref{fig4}c are extracted from linear fits of the rf frequency dependent FMR linewidth obtained from the $S_{11}$ measurement (see Supplementary Fig.$\,$S2).

\noindent\textbf{Rf-power reflection and spin pumping.} We compare the nonlocal results with the microwave reflection and spin-pumping measurement in Fig.$\,$\ref{dispersion_for_three_fields_two_thicknesses}, where the rf field oscillates at a frequency of 3$\,$GHz. The spin-pumping voltage is sensitive to processes near the surface \cite{jungfleisch2015thickness}, whereas the microwave power reflection is a method to probe the global magnetization.

At low rf power, when we increase the static field, the reflected microwave power (see Fig.$\,$\ref{dispersion_for_three_fields_two_thicknesses}b) starts to decrease at the static field of $\sim\pm33\,$mT, where we see a prominent change of the nonlocal signal in Fig.$\,$\ref{dispersion_for_three_fields_two_thicknesses}a. When the static field reaches $\pm54\,$mT, a sharp dip appears for the reflected rf power with similar amplitude and lineshape at both positive and negative static fields, which is a result of microwave power absorption by YIG at FMR. At static fields lower than the resonance condition, the absorption takes place due to the available perpendicular standing spin wave modes \cite{jungfleisch2015thickness}. By comparison, the measured spin-pumping voltage changes sign when the static field changes polarity. At low rf power in Fig.$\,$\ref{dispersion_for_three_fields_two_thicknesses}c, it also has a big shoulder at fields lower than FMR condition, starting at the static field of $\sim33\,$mT. However, it shows larger amplitude at positive resonance field than that at the negative one.

\begin{figure*}[tb]
	\centering
	\includegraphics[width=0.65\linewidth]{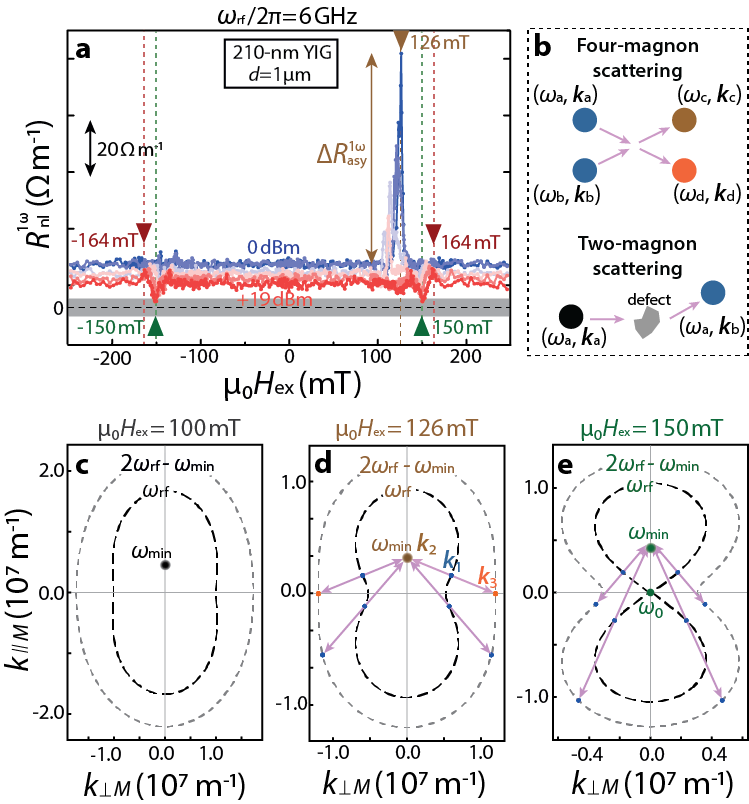}
	\caption{\textbf{Comparing nonlocal resistance with relevant iso-frequency curves to illustrate the kinetic process.} \textbf{a} First harmonic nonlocal results with different applied rf powers, from 0$\,$dBm (blue) to $+19\,$dBm (red). The scale bar is 20$\,\Omega\,$m$^{-1}$. The brown, green and red triangles with corresponding vertical dashed lines indicate special field values. \textbf{b} Magnon scattering processes. Four-magnon scattering: A magnon with frequency of $\omega_{\textrm{a}}$ and momentum of $\textbf{\textit{k}}_{\textrm{a}}$, denoted as ($\omega_{\textrm{a}}$, $\textbf{\textit{k}}_{\textrm{a}}$), scatter with the other magnon ($\omega_{\textrm{b}}$, $\textbf{\textit{k}}_{\textrm{b}}$), giving rise to two magnons ($\omega_{\textrm{c}}$, $\textbf{\textit{k}}_{\textrm{c}}$) and ($\omega_{\textrm{d}}$, $\textbf{\textit{k}}_{\textrm{d}}$), where energy and momentum are conserved ($\omega_{\textrm{a}}+\omega_{\textrm{b}}=\omega_{\textrm{c}}+\omega_{\textrm{d}}$, $\textbf{\textit{k}}_{\textrm{a}}+\textbf{\textit{k}}_{\textrm{b}}=\textbf{\textit{k}}_{\textrm{c}}+\textbf{\textit{k}}_{\textrm{d}}$). Two-magnon scattering: One incoming magnon ($\omega_{\textrm{a}}$, $\textbf{\textit{k}}_{\textrm{a}}$) scatters with a defect in the system, resulting in an outcoming magnon ($\omega_{\textrm{a}}$, $\textbf{\textit{k}}_{\textrm{b}}$), where energy is conserved but momentum is not. \textbf{c}-\textbf{e} Isofrequency curves of 210$\,$nm thick YIG dispersion relation at static fields of 100$\,$mT, 126$\,$mT and 150$\,$mT. $k_{\perp M}$ and $k_{\parallel M}$ are wavevectors perpendicular and parallel to the in-plane magnetization, corresponding to the magnetostatic surface mode and backward volume mode. Magnon frequencies of $\omega_{\textrm{rf}}$ and $2\omega_{\textrm{rf}}-\omega_{\textrm{min}}$ are in black and gray dashed lines, respectively. $\omega_{\textrm{rf}}/2\pi$ is $6\,$GHz and $\omega_{\textrm{min}}$ is the band minimum at corresponding static field. In \textbf{d} and \textbf{e}, blue dots on the iso-frequency lines of $\omega_{\textrm{rf}}$ represent the initial states of the four-magnon scattering. Orange or blue dots on the iso-frequency line of $2\omega_{\textrm{min}}-\omega_{\textrm{rf}}$ represent one of the final states of the four-magnon scattering. The orange one has large $k_{\parallel M}$ component and group velocity in the same direction. We draw the lowest magnon dispersion relation with parameters obtained from the Kittel fit of rf power reflection measurement: Gyromagnetic ratio $\gamma=27.3\,$GHz/T and saturation magnetization $M_{\textrm{s}}=135\,$kA/m. We used exchange stiffness of $1\times10^{-39}\,$J$\,$m$^{2}$ \cite{kalinikos1986theory}.
	}
	\label{isofrequencycurves}
\end{figure*}

At high rf power, the reflected microwave power dips at FMR in Fig.$\,$\ref{dispersion_for_three_fields_two_thicknesses}d have larger linewidth, and they do not show the big shoulder at fields lower than FMR condition, compared with the low rf power case. The spin-pumping voltages in Fig.$\,$\ref{dispersion_for_three_fields_two_thicknesses}e show similar amplitudes and lineshapes at positive and negative static fields. The highly distorted lineshape at high rf power is a result of the nonlinear FMR \cite{gui2009foldover,gui2009direct}. Correspondingly, the nonlocal signals at both positive and negative static field show a suppression at the FMR and an enhancement at a field higher than the FMR.

The resonance fields for different rf frequencies and samples are confirmed with the $S_{11}$ (see Supplementary Fig.$\,$S2) and spin-pumping measurements.

\begin{flushleft}
	\textbf{DISCUSSION}
\end{flushleft}

Inspired by a recent work \cite{kreil2018kinetic}, we compare the first harmonic nonlocal result and the isofrequency lines of the YIG dispersion relation as shown in Fig.$\,$\ref{isofrequencycurves}. We find that at relevant static fields, where the distinct changes of nonlocal resistances appear, various kinetic processes in the magnon cloud are allowed according to energy and momentum conservation. This changes the magnon scattering and the occupation of relevant magnon states so as to alter the transport.

Firstly, at a static field lower than the FMR condition, $\sim126\,$mT, we observe a strong increase of $\textit{R}_{\textrm{nl}}^{1\omega}$ in Fig.$\,$\ref{isofrequencycurves}a. The applied rf power is relatively low, $\sim1\,$mW. At this static field, magnons with frequency of $\omega_{\textrm{rf}}$ and momentum of $\textbf{\textit{k}}_{1}$, denoted as ($\omega_{\textrm{rf}}$, $\textbf{\textit{k}}_{1}$), can efficiently scatter with each other as shown in Fig.$\,$\ref{isofrequencycurves}d. This results in one magnon at the band minimum ($\omega_{\textrm{min}}$, $\textbf{\textit{k}}_{2}$) and the other with higher frequency ($2\omega_{\textrm{rf}}-\omega_{\textrm{min}}$, $\textbf{\textit{k}}_{3}$), obeying energy and momentum conservation for the four-magnon scattering process ($2\textbf{\textit{k}}_{1}=\textbf{\textit{k}}_{2}+\textbf{\textit{k}}_{3}$), as illustrated in Fig.$\,$\ref{isofrequencycurves}b. The resulting high energy $\textbf{\textit{k}}_{3}$-magnons possess large $k_{\perp M}$ component, which is characteristic of the chiral surface mode \cite{eshbach1960surface}. Besides, they have a group velocity pointing in the direction of $k_{\perp M}$. There are various broadening processes, such as two-magnon scattering (see Fig.$\,$\ref{isofrequencycurves}b) and phonon-related Gilbert damping, that prepare the magnons with momentum $\textit{\textbf{k}}_{1}$. To compare, we also show the isofrequency lines at 100$\,$mT in Fig.$\,$\ref{isofrequencycurves}c, where there is no effective kinetic process in the manner described above. 

The chiral property of the surface mode excited by the kinetic process might explain why the strong enhancement of nonlocal signals only appears at the positive static field of 126$\,$mT. 
Moreover, the nonreciprocal property of the surface mode \cite{an2013unidirectional,jungfleisch2015thickness} manifests itself as shown in Fig.$\,$\ref{fig4}f, where this asymmetric feature is altered upon interchanging the roles of injector and detector. Besides, the asymmetric feature is much more significant in the 210$\,$nm thick YIG compared to the 100$\,$nm thick one (see Supplementary Fig.$\,$S8), because the number of standing spin wave modes increases with increasing film thickness, which enhances the scattering probability \cite{jungfleisch2015thickness}. In addition, this corroborates the chiral surface mode origin, because magnetostatic surface modes are better localized at the top and bottom surface in the thicker YIG. Also, $\Delta R_{\textrm{asy}}^{1\omega}$ increases drastically with rf power until $\sim10\,$mW, where it starts to decrease. The drastic enhancement of the nonlocal signals might be related with the increasing occupation of the magnon band minimum due to the kinetic process, which significantly facilitates the magnon conduction. Heating due to rf power might be the reason why we do not see this effect at higher rf power \cite{bozhko2016supercurrent}. Last but not least, this can also explain the different spin-pumping voltages for the positive and negative static fields at relatively low rf power as shown in Fig.$\,$\ref{dispersion_for_three_fields_two_thicknesses}c, because the spin-pumping voltage is sensitive to processes near the surface \cite{jungfleisch2015thickness}. Therefore, the secondary magnons with chiral surface mode character contribute to the spin-pumping voltage differently for positive and negative static fields.

Secondly, at a static field corresponding to the FMR condition, $\sim150\,$mT, we observe a suppression of the nonlocal signals at relatively large rf power in Fig.$\,$\ref{isofrequencycurves}a. As shown in Fig.$\,$\ref{isofrequencycurves}e, the frequency of the FMR mode coincides with the rf frequency ($\omega_{\textrm{rf}}=\omega_{0}$). Heating due to microwave absorption at FMR opens a bigger precession cone angle and reduces the effective magnetization, which can reduce nonlocal signals (see METHOD). However, this can not explain a suppression as large as 95$\%$.

The saturating trend of $\Delta R_{\textrm{sup}}^{1\omega}$ with rf power as shown in Fig.$\,$\ref{fig4}b is similar to that of the damping parameter in the inset of Fig.$\,$\ref{fig4}c. This suggests a highly nonlinear FMR (NFMR) regime \cite{dobin2003intrinsic,anderson1955instability,suhl1956nonlinear,suhl1957theory}, where a FMR mode quickly transfers its energy to other degenerate non-zero momentum modes, as indicated by the blue dots on the isofrequency curves of $\omega_{\textrm{rf}}$ in Fig.$\,$\ref{isofrequencycurves}e, which also experience the 4-magnon scattering. This process redistributes the energy, which is also known as second-order Suhl spin wave instability \cite{sparks1964ferromagnetic}. Based on this, one possible reason for the strong suppression of the nonlocal signals at the resonance is that the electrically injected magnons are strongly scattered off by the secondary magnons due to NFMR,  but we do not have a full understanding. 

Depending on the strength of the rf field, the magnon system reacts differently: With increasing $h_{\textrm{rf}}$, more magnons are generated, which means a larger deviation from equilibrium. As soon as the deviation surpasses a critical point, the magnon-magnon interaction starts to determine the behavior of the magnon system and the system switches from the linear to the nonlinear regime \cite{l1990nonlinear,lavrinenko1981kinetic,boardman1988three}. The threshold rf field separating linear and nonlinear regimes is on the order of 0.01$\,$mT for a single crystal YIG film \cite{suhl1957theory, suhl1956nonlinear,anderson1955instability,cottam1994linear}, which is easily achievable due to the low damping of YIG. The regime of rf field we use is approximately from $0.03\,$mT to $1\,$mT, which is above the threshold rf field. Therefore, we observe the influence of the microwave field on nonlocal signals for the in-plane magnetization in the nonlinear regime.

Thirdly, at a static field higher than the FMR condition, $\sim164\,$mT, we observe an increase of $\textit{R}_{\textrm{nl}}^{1\omega}$ in Fig.$\,$\ref{isofrequencycurves}a.
At this static field, the rf frequency coincides with the band minimum ($\omega_{\textrm{rf}}=\omega_{\textrm{min}}$). Rf-power pumps magnons at the band minima, where there is a large amount of available states. Even though they have zero group velocity, sufficient increase of the magnon numbers enhances the overall magnon conductivity \cite{cornelissen2018spin}, giving rise to an enhancement of nonlocal signals.

The three relations discussed above, between the field position of the distinct change of nonlocal resistances and the special position of the dispersion relation with respect to the rf frequency, also apply to 3$\,$GHz and 9$\,$GHz rf frequency and 100$\,$nm YIG cases, which can be found in the Supplementary Fig.$\,$S8.

To conclude, we observe that an rf field can strongly influence magnon spin transport for an in-plane magnetization via kinetic processes and the occupation of relevant magnon states. Firstly, with relatively small rf power, we observe a strong enhancement of the nonlocal signal as large as 800$\%$. It appears only at a positive static field smaller than the FMR condition. This might be a result of effective kinetic processes: the increasing occupation of the magnon band minimum facilitates magnon conduction, and the resulting chiral surface mode manifests itself differently in the positive and negative static fields. Secondly, with large rf power, nonlinear FMR triggers a strong suppression of incoherent thermal magnon transport. This phenomenon is partially due to the heating from microwave power, but the main contribution still remains unclear. We assume that it is related with the scattering between secondary magnons from the NFMR and thermal magnons, which needs more theoretical confirmation. Thirdly, at a static field slightly higher than FMR an enhancement of the thermal magnon transport has been observed, when the rf frequency coincides with the frequency of magnons at the band minima, where the most states are available. This increases the total magnon number resulting in an enhancement of the overall magnon spin conductivity. Therefore, our results show that microwave fields can effectively control thermal magnon spin transport. Moreover, nonlocal magnon spin transport can be used to perform spectroscopy of magnons or spin waves. Our results present a new opportunity to couple electromagnetic signals with spin waves, providing a way to scale down microwave electronics operating at GHz frequencies.

\begin{flushleft}
	{\textbf{ACKNOWLEDGEMENT}}
\end{flushleft} We appreciate G. E. W. Bauer for suggestions and we thank T. Kuschel, T. van der Sar, P. S. Haughian, M. Weiler, A. Kamra, J. Shan, K. Oyanagi and M. Schabes for discussions. We acknowledge H. M. de Roosz, J. G. Holstein, H. Adema and T. J. Schouten for their technical assistance. This work is part of the research program Magnon Spintronics (MSP) No.$\,$159 financed by the Netherlands Organisation for Scientific Research (NWO). We appreciate support from the NanoLab NL and the Zernike Institute for Advanced Materials. This research is partly financed by the NWO Spinoza prize awarded to Prof. B. J. van Wees. Further support by EU FP7 ICT Grant No.$\,$612759 InSpin and Marie Curie initial training network Spinograph (607904) is gratefully acknowledged.

\begin{flushleft}
	\textbf{AUTHOR CONTRIBUTIONS}
\end{flushleft} J.L. and B.J.W. initiated the project. J.L. and F.F. performed the experiment and analyzed the data with support from L.J.C. and J.C.L. The theoretical work was developed by J.L. with contributions from F.F. and support from B.F., R.D. and B.J.W. The manuscript was written by J.L. with contributions from all the coauthors.

\begin{flushleft}
	\textbf{ADDITIONAL INFORMATION}
\end{flushleft} 

\begin{flushleft}
	\textbf{Competing interests:} The authors declare no competing interests.
\end{flushleft}

\bibliography{reference}

\begin{flushleft}
	\small{\textbf{METHODS}}
\end{flushleft}
\small{\textbf{Sample preparation.} The 210$\,$nm and 100$\,$nm thick single crystal yttrium iron garnet (YIG, Y$_{3}$Fe$_{5}$O$_{12}$) film was grown by liquid phase epitaxy (LPE) on top of a 500$\,\mu$m thick single crystal (111) gadolinium gallium garnet (GGG, Gd$_{3}$Ga$_{5}$O$_{12}$) substrate, commercially obtained from the company Matesy GmbH. The saturation magnetization for 210$\,$nm and 100$\,$nm YIG sample are $174\pm4\,$mT and $173\pm2\,$mT, respectively. The Gilbert damping parameters for both samples are on the order of $1\times10^{-4}$. The devices were patterned by electron beam lithography: 7-nm-thick Pt strips were deposited on YIG by dc sputtering. After that Ti$|$Au layers of 5$|$75$\,$nm were deposited by e-beam evaporation, which includes Ti$|$Au leads contacted to Pt strips and an on-chip stripline with a shorted end connected to a vector network analyser (VNA).\hfill \break}

\noindent\small{\textbf{Experimental setup.} 
\small{The sample is positioned between a pair of magnetic poles, giving rise to an external static field $\textbf{\emph{H}}_{\textrm{ex}}$, which aligns the magnetization of the YIG in the film plane, perpendicular to the Pt strips. The microwave power is provided by a vector network analyzer (VNA, Rohde $\&$ Schwarz ZVA-40). By connecting VNA to the on-chip stripline rf field $\textit{\textbf{h}}_{\textrm{rf}}$ is generated with largest amplitude at the shorted end of the stripline. For the connection, a picoprobe (type 40A-GS-400-LP) and SMA-connectors are used. In order to be able to use lockin detection for the spin pumping measurement with small signals, the rf field is modulated either by the rf frequencies $\omega_{\textrm{rf}}$ or rf powers $P_{\textrm{rf}}$ with a lockin frequency $\omega_{\textrm{lockin}}$ of a few Hz. In order to use $\omega_{\textrm{lockin}}$ to modulate the rf signal, we use a frequency-doubler. Since the VNA is only triggered by upgoing (or downgoing) edges of an input pulse, the outcoming signal has a frequency of half of the frequency of the triggering signals. Therefore, with a frequency doubler we can equalize the modulating frequency of the outcoming rf signal with $\omega_{\textrm{lockin}}$.\hfill \break}

\noindent\small{\textbf{Local measurement.} 
	\small{Compared with nonlocal measurements, local refers to the experiment where we send a current through a Pt strip and measure the voltage across the same strip. For this, we used one of the Pt strips of the nonlocal devices. The results can be found in Supplementary Fig.$\,$S4. Purposes of this experiment are two-fold: Firstly, by comparing the local and nonlocal measurement we can discuss the influence of rf field on the injector and detector Pt strips, which is found out to be marginal compared with the influence of rf field on the magnon spin transport. Secondly, Pt can be used as a resistive temperature sensor since its resistance scales linearly with the temperature. 
		
	By increasing the rf power the local resistance enhances due to the heating (see Supplementary Fig.$\,$S4a). From the lowest to highest applied rf power, the resistance increases by $\sim0.13\,$k$\Omega$ (see Supplementary Fig.$\,$S4b). Since the Pt strip has the smallest cross-section comparing with the Ti$|$Au leads, it is the most resistive part. Therefore, we assume the resistance increase is mainly due to Pt. This corresponds to a resistivity increase of $\sim1.5\times10^{-8}\,\Omega\,$m and a maximum temperature enhancement of $\sim35\,$K, according to
		\begin{equation}
		\Delta\rho_=C\Delta T
		\end{equation}
		where $\Delta\rho$ is the resistivity change, $\Delta$T is the temperature change and $C$ is the linear scaling factor of $4.3\times10^{-10}\,\Omega\,$mK$^{-1}$, obtained from the temperature dependent resistivity measurement for the same type of Pt thin film near room temperature. The possible causes of the temperature increase are as following: Firstly, Joule heating is accompanied with the rf current passing through the on-chip stripline, which is not far away from the Pt devices, around $10\,\mu$m. Secondly, the eddy currents of all conducting leads including Pt and Ti$|$Au can produce additional heat. Thirdly, since the Ti$|$Au leads are located even closer to the stripline, under a temperature gradient additional current due to conventional Seebeck effect can be generated accompanied by the Joule heating. Fourthly, due to the Peltier effect the connection between Pt and Ti$|$Au can heat up or cool down.

		There is a small decrease of the Pt resistance of $\sim2.7\,\Omega$ at the FMR condition, which translates to a Pt resistivity change of $3.2\times10^{-10}\,\Omega\,$m and a temperature change of 0.74$\,$K (see Supplementary Fig.$\,$S4c). This is a result of the fact that at FMR more rf power is delivered to magnetic dynamics instead of increasing the lattice temperature, which results in a lower resistance for the Pt strip. Besides, we can completely exclude the influence from the ac spin-pumping effect on the quasi-dc lockin measured resistance, because for both positive and negative static fields we see a relative decrease of the resistance; however, opposite signs are expected for the spin-pumping signals by changing the polarity of the static field. Most importantly, this Pt resistance change of $\sim2.7\,\Omega$ at the FMR condition under the highest applied rf power is still negligible compared with the resistance of the Pt strip, which is around 1.4$\,$k$\Omega$; therefore, the change of the Pt resistance can not explain the strong change of the nonlocal signals at the FMR condition.\hfill \break}

\noindent\small{\textbf{Comparison between the rf power dependent $R^{1\omega}_{\textrm{nl}}$ and $R^{2\omega}_{\textrm{nl}}$.} 
	\small{We compared the first and second harmonic signals measured simultaneously at different applied rf powers, from $-10\,$dBm to $+19\,$dBm (see Supplementary Fig.$\,$S1). At relatively low rf power, both first and second harmonic signals show features appearing only at the positive static field. With increasing rf power, these asymmetric features initially become more pronounced, until a certain applied power of $\sim+15\,$dBm they start to decrease and eventually vanish. Before vanishing, the suppression and enhancement of the nonlocal signals near the FMR condition and at a higher static field than FMR condition arise at both positive and negative static field. Qualitatively, the results of the first and second harmonic signals are consistent. However, the asymmetric feature is more significant in the first harmonic signals than that of the second harmonic signals. This quantitative difference is probably due to different generation mechanism of magnon accumulation by the electron magnon spin injection for the first harmonic signals and by the temperature gradient for the second harmonic signals. \hfill \break}

\noindent\small{\textbf{Rf-power reflection measurement.} 
\small{In order to characterize the magnetization dynamics of our sample, we performed standard FMR resonance, where the rf field $h_{\textrm{rf}}$ is mostly perpendicular to the external field (in-plane) at the device. To detect the FMR signals, one way is to probe the reflected rf power, i.e. $S_{11}$-parameter. Both high and low rf powers are applied as shown in Fig.$\,$\ref{dispersion_for_three_fields_two_thicknesses}b and \ref{dispersion_for_three_fields_two_thicknesses}d. We analyzed the results of the microwave reflection measurement: Firstly, we fit the static field dependent resonance frequencies from the $S_{11}$ measurement based on Kittel equation \cite{kittel1948theory} (see Supplementary Fig.$\,$S2a), confirming that the excited mode at various rf powers is the uniform precession mode, i.e. FMR. We also observed that due to the microwave heating, the largest upshifting of the $S_{11}$ peak position is $\sim$4$\,$mT at 9$\,$GHz. Secondly, we extracted the linewidth of the $S_{11}$ peaks, i.e. the full width at half maximum (FWHM), as a function of the rf frequencies at different rf powers (see Supplementary Fig.$\,$S2b), from which we obtained the effective Gilbert damping parameter $\alpha$ by the linear fitting based on \cite{kalarickal2006ferromagnetic}
\begin{equation}
\Delta H=\frac{4\pi\alpha}{\mu_{0}\gamma}f_{\textrm{rf}}+\Delta H_{0}
\end{equation} where $\mu_{0}\Delta H$ corresponds to the full width at half maximum (FWHM) of the $S_{11}$ peaks with $\mu_{0}$ being the vacuum permeability, $\Delta H_{0}$ is related to the extrinsic damping caused by inhomogeneities and impurities, $\gamma$ is the gyromagnetic ratio and $f_{\textrm{rf}}$ and $\omega_{\textrm{rf}}$ are the frequency and the angular frequency of the rf field ($f_{\textrm{rf}}=\omega_{\textrm{rf}}/2\pi$). In Supplementary Fig.$\,$S2c and the inset of Fig.$\,$\ref{fig4}c we show the effective Gilbert damping parameter $\alpha$ as a function of rf power. The nonlinear FMR is further confirmed by the saturating behavior of $\alpha$ as a function of rf power shown in Supplementary Fig.$\,$S2c and the inset of Fig.$\,$\ref{fig4}c \cite{gui2009foldover,gui2009direct}.\hfill \break}

\noindent\small{\textbf{Spin-pumping measurement.} 
\small{The other method to detect the FMR signals is to measure the spin-pumping signals ($V_{\textrm{sp}}$), i.e. the ISHE voltage transversally to the external field by one of the Pt strips used for the magnon injection and detection. Since $V_{\textrm{sp}}$ scales with the length of the Pt strip and in our case it is only 12$\,\mu$m, we expect $V_{\textrm{sp}}$ to be small. In order to still be able to detect it, we have to use the lockin technique to modulate between two rf powers or rf frequencies. In Fig.$\,$\ref{dispersion_for_three_fields_two_thicknesses}c and \ref{dispersion_for_three_fields_two_thicknesses}e, we used a low lockin frequency of 7.777$\,$Hz to modulate between two rf powers of $P^{1}_{\textrm{rf}}$ and $P^{2}_{\textrm{rf}}$ with rf frequency of 3$\,$GHz. \hfill \break}

\noindent\small{\textbf{Excitation current dependency.} 
	\small{We performed nonlocal measurement with various excitation current with rms-amplitudes of 100$\,\mu$A, 150$\,\mu$A, 200$\,\mu$A and 250$\,\mu$A. The typical one we use is 200$\,\mu$A. With higher rf power, the suppression and enhancement at FMR and at a higher field than FMR scale with the excitation current (see Supplementary Fig.$\,$S6a and S6b). At lower rf power the strong enhancement happening only on the positive static field also scales with the excitation current (see Supplementary Fig.$\,$S6c). Therefore, we conclude that all the processes including the magnon injection, detection and transport are the linear responses of the system to the excitation current.\hfill \break}

\noindent\small{\textbf{Onsager reciprocity.} 
	\small{In the linear-response regime, Onsager reciprocity holds \cite{PhysRev.37.405, PhysRev.38.2265, jacquod2012onsager}. This means that the transport coefficient matrix equals its transposed counterpart, which represents the time-reversed system, e.g. reversing the external field. In our nonlocal results, we generally have the first harmonic signals being evenly symmetric for positive and negative static field due to the reciprocal properties of the linear transport \cite{cornelissen2015long}. When we interchange the role of injector and detector, the signals remain the same because the injector and detector are designed identically. 
		
	From the static field dependent results in Fig.$\,$\ref{dispersion_for_three_fields_two_thicknesses}a and \ref{isofrequencycurves}a, we already see that the features at the FMR and above the FMR condition are mostly symmetric for positive and negative field, which indicates the reciprocity holds. However, this is not the case for the feature at a static field lower than the FMR condition, which only appears at positive field. Also, we interchange the role of injector and detector. This has been conducted at different rf powers (see Supplementary Fig.$\,$S7). Indeed, signals remain the same for the suppression at the FMR condition and enhancement at a higher field than FMR as pointed out by the well overlapping between the results of IV and VI configurations at relatively high rf power (see Supplementary Fig.$\,$S7b and S7c). However, signals change significantly for the asymmetric features which only appear at the positive field at relatively low rf power (see Supplementary Fig.$\,$S7a and S7b), which is a result of the non-reciprocal magnetostatic surface mode \cite{camley1987nonreciprocal}. We think this is related with the secondary magnons with chiral surface mode character; however, we do not completely understand the mechanism of the interplay between the chiral mode and the electrically injected thermal magnon spin current. \hfill \break}

\noindent\small{\textbf{Rf-power calibration.} 
	\small{In order to be able to transfer the maximum amount of power from the VNA to the on-chip stripline, we design the stripline in such a way that its impedance matches the VNA impedance optimally for the operating rf frequencies (under 10$\,$GHz). The delivered rf power $P_{\textrm{rf}}^{\textrm{c}}$ shown in Fig.$\,$\ref{fig4}a-\ref{fig4}e is calibrated in the following way: We measured the load impedance as 28.9$\,\Omega$, 39.8$\,\Omega$ and 53.5$\,\Omega$ by the VNA at rf frequencies of 3$\,$GHz, 6$\,$GHz and 9$\,$GHz, respectively. When the load impedance is close to 50$\,\Omega$, which is the source impedance, the maximum rf power is transferred. Therefore, with the same applied rf power the 9$\,$GHz rf frequency one with impedance of 53.5$\,\Omega$ has the highest delivered rf power.  \hfill \break}

\noindent\small{\textbf{Temperature effect due to the rf power.} 
	\small{Rf-power heats up the system in a nontrivial way \cite{iguchi2016measurement}. As a result, the temperature profile changes: At a certain position of the devices, both the base temperature (lattice temperature and magnon temperature) and the corresponding temperature gradient change. For the device we used, the change of temperature caused by rf power has following effects:
		
		From Pt resistance measurement, we know that the maximum lattice temperature increase caused by the microwave power is $\sim35\,$K. Further, we study the change of the resonance peak position by Kittel equation from Supplementary Fig.$\,$S2a, the maximum change of the effective magnetization is 3$\,$mT, which corresponds to 7.5$\,$K temperature increase at FMR, if we assume linear temperature dependency of the saturation magnetization of YIG near room temperature with a conversion factor of 0.4$\,$mT/K \cite{hansen1974saturation}. Therefore, the crystal lattice heats up more than the spin lattice. 
		
		Besides, in the nonlocal results the position where we see the suppression also shifts towards the higher field with increasing the rf power as shown in Fig.$\,$\ref{dispersion_for_three_fields_two_thicknesses}. The maximum shift is $\sim10\,$mT, which is larger than the shifting of the resonance condition measured by $S_{11}$ parameter, maximum 4$\,$mT. However, at high rf power the position of the resonance measured by $S_{11}$ parameter and the position of the strong suppression in the nonlocal signals are the same as can be seen in Fig.$\,$\ref{dispersion_for_three_fields_two_thicknesses}a and \ref{dispersion_for_three_fields_two_thicknesses}d. Therefore, the position shift of the suppression feature in the nonlocal results at lower rf power cannot be fully explained by the heating due to microwave power at FMR. There are other effects presenting in the nonlocal measurements. 
		
		Moreover, we observe a shift of the background or offset for the first and second harmonic nonlocal signals as shown in Supplementary Fig.$\,$S5a and S5c, respectively. We see that both the change of the first and second harmonic signal offsets, $\Delta R^{1\omega}_{\textrm{offset}}$ and $\Delta R^{2\omega}_{\textrm{offset}}$, scale linearly with the change of the measured local resistance, $\Delta R_{\textrm{local}}$ (see Supplementary Fig.$\,$S5a-S5d). The possible reason is that Pt and Ti$|$Au leads experience a conventional Peltier and Seebeck effect, which can be detected by the lockin voltage detector. The measured voltages scales with the resistance of the metal leads, which increases with rf power as shown in Supplementary Fig.$\,$S4b. Besides, the Seebeck coefficient can change with temperature, which will complicate the situation even more. Generally if the Ti$|$Au strips are more symmetrically distributed with respect to the Pt strip, we do not encounter this offset or less of the second harmonics signals. From device to device, the offset also varies because the designs are not completely the same. We can not completely resolve the origin of the offset shifting, but it is likely due to the thermal effect in the devices.

		Also, with increasing the rf power the magnitude of the second harmonic signals also changes, which can be either increase or decrease (see Supplementary Fig.$\,$S5e). This is a result of the change of Pt-current induced temperature gradient due to the change of rf power, which varies from device to device.
		
		 \hfill \break}

\noindent\small{\textbf{Rf-field strength and precession cone angle.} 
	\small{We discuss the precession cone angle $\theta$, since cone angle opening might influence the injection and detection efficiencies, especially at the resonance condition with large $\theta$. Therefore, it might affect the results of the nonlocal resistance. With increasing the rf power, the present $h_{\textrm{rf}}$ at the device enhances, which results in an increasing $\theta$ at the FMR mode. Until reaching a certain rf field for the given magnet, $\theta$ will not increase any more. This is called the Suhl saturation \cite{suhl1957theory, suhl1956nonlinear, HAGHSHENASFARD2017380}.  
		
		First we estimate the strength of the rf field ($h_{\textrm{rf}}$) at the device by applying the Biot-Savart law. For simplicity, we use the infinite wire assumption so that
		\begin{equation}
		h_{\textrm{rf}}=\frac{I_{\textrm{rf}}}{2\pi r}
		\end{equation}
		where $r$ is the distance between the stripline and the middle of the Pt strips, which is $\sim10\,\mu$m. $I_{\textrm{rf}}$ is the rms-value of the radio-frequency oscillating current provided by VNA, which is calculated using
		\begin{equation}
		I_{\textrm{rf}}=\sqrt{P_{\textrm{rf}}/R}
		\end{equation}
		where $P_{\textrm{rf}}$ is the rf power provided by VNA and $R$ is the impedance of the stripline, which is measured by VNA. $\theta$ can be estimated by \cite{guan2007phase} 
		\begin{equation}
		\theta=\frac{h_{\textrm{rf}}}{\Delta H}
		\label{thelta}\end{equation}
		where $\Delta H$ is the linewidth of the reflection peaks from the $S_{11}$ measurement at corresponding rf power.
		
		Injection and detection efficiencies are maximum when the electron spin polarization is parallel or anti-parallel to the net magnetization of YIG. However, when $\theta$ increases with a GHz precession frequency, the electron spin polarization has less component parallel to the effective YIG magnetization, which results in the decrease of the injection and detection efficiency. The first and second harmonic nonlocal resistances scale with $\cos\theta$ and $\cos^{2}\theta$, respectively. 
		
		According to Eq.$\,$\ref{thelta}, the estimated maximum rf field is $\sim1\,$mT, corresponding to $\mu_{0}\Delta H$ being $\sim4\,$mT for the highest rf power (see Supplementary Fig.$\,$S2b). This gives rise to a precession cone angle of 15$^{\circ}$, which results in a reduction of 7$\%$ and 4$\%$ for the first and second harmonic signals, respectively. However, we see a much larger reduction of more than 70$\%$ of the nonlocal first harmonic signals at the highest applied rf power. Therefore, we can conclude that the increase of the precession cone angles is not the cause of strong suppression of the nonlocal signals at FMR condition. \hfill \break}

\noindent\small{\textbf{Comparison of 210$\,$nm and 100$\,$nm thick YIG results.} 
	\small{We compare the first harmonic nonlocal results of 210$\,$nm and 100$\,$nm thick YIG under an rf field oscillating at 3$\,$GHz, 6$\,$GHz and 9$\,$GHz (see Supplementary Fig.$\,$S8). The results for 210$\,$nm YIG at rf frequency of 3$\,$GHz and 6$\,$GHz are shown in Fig.$\,$\ref{dispersion_for_three_fields_two_thicknesses}a and \ref{isofrequencycurves}a, respectively. Since the properties of the magnon dispersion relation depend on the thickness of YIG, we studied the relevant iso-frequency lines of the dispersion relation for 210$\,$nm and 100$\,$nm thick YIG at special static fields. Besides, the possible four-magnon scattering illustration is provided.
		
		The asymmetric feature is much more significant in the 210$\,$nm thick YIG, even though the four-magnon scattering process is also possible for 100$\,$nm thick YIG (see Fig.$\,$S8e, S8k and S8q). There are two possible reasons: Firstly, the thinner the YIG film, the less dense will be the dispersion relation, which can provide less available states for the two magnon scattering to prepare the initial states of the 4-magnon scattering \cite{jungfleisch2015thickness}. Secondly, it corroborates the chiral surface mode origin, because magnetostatic surface modes are better localized at the top and bottom surface in the thicker YIG.

		Moreover, due to the larger difference between the surface and bulk mode frequencies with the same momentum for 210$\,$nm thick YIG, we see static field corresponding to the enhancement at a higher field than FMR is larger in 210$\,$nm YIG (164$\,$mT in Supplementary Fig.$\,$S8g) than that in 100-nm thick YIG (155$\,$mT in Supplementary Fig.$\,$S8j) for the case of 6$\,$GHz rf frequency. This relation also applies to 3$\,$GHz and 9$\,$GHz rf frequency cases. \hfill \break}

 \noindent\small{\textbf{Extracting $\Delta R_{\textrm{enh}}^{1\omega}$, $\Delta R_{\textrm{enh}}^{2\omega}$, $\Delta R_{\textrm{sup}}^{1\omega}$, $\Delta R_{\textrm{sup}}^{2\omega}$ and $\Delta R_{\textrm{asy}}^{1\omega}$.} 
 	\small{We defined the suppression at the FMR condition and the enhancement at a field higher than the FMR condition for the first and second harmonic nonlocal signals in Fig.$\,$\ref{fig1}b ($\Delta R^{1\omega}_{\textrm{sup}}$, $\Delta R^{1\omega}_{\textrm{enh}}$) and Fig.$\,$\ref{fig1}c ($\Delta R^{2\omega}_{\textrm{sup}}$, $\Delta R^{2\omega}_{\textrm{enh}}$), respectively. The enhancement of the first harmonic signals at a positive static field smaller than the FMR condition is defined as $\Delta R^{1\omega}_{\textrm{asy}}$ in Fig.$\,$\ref{isofrequencycurves}a. We studied their rf power dependency in Fig.$\,$\ref{fig4}a-\ref{fig4}e.  To extract the amplitude of them, a consistent method has been employed as following for results at different rf power and rf frequency. 
 		
 	We denoted the amplitude of the first and second harmonic signals as $F_{\textrm{n}}$ and $S_{\textrm{n}}$, respectively. Specifically, n=0 corresponds to the baseline resistance, n=1,2,3 represent the magnitudes of the nonlocal resistances at the special static fields (see Supplementary Fig.$\,$S3a and S3b). $+$ and $-$ are used to identify the features at positive and negative static fields, respectively. For n=1, it is the feature only appears at the positive static field. Since this feature is much more significant in $R^{1\omega}_{\textrm{nl}}$, we only discuss it in the first harmonic nonlocal resistances and its magnitude is $\Delta R^{1\omega}_{\textrm{asy}}$.
 	
 	The amplitude of the first and second harmonic nonlocal resistances are defined as
 	\begin{equation}
 	R^{1\omega}_{\textrm{nl}}=|F_{0}|
 	\end{equation}
 	\begin{equation}
 	R^{2\omega}_{\textrm{nl}}=\frac{1}{2}\,|S_{0-}-S_{0+}|
 	\end{equation}
 	where baseline resistances $F_{0}$, $S_{0-}$ and $S_{0+}$ are the average of the data from a region, which is below the FMR condition without features due to the rf field (see purple areas in Supplementary Fig.$\,$S3a and S3b). Specifically, we took them from $-0.3\,\mu_{0}H_{\textrm{FMR}}$ to $-10.0\,$mT or from $10.0\,$mT to $0.3\,\mu_{0}H_{\textrm{FMR}}$, where $\mu_{0}H_{\textrm{FMR}}$ is the field corresponding to the FMR condition. For the first harmonic signals, only the mean value at the positive field is taken as $F_{0}$, while for the second harmonic signals the mean values are extracted at both negative and positive field as $S_{0-}$ and $S_{0+}$.  
 	
 	With higher rf power, when the nonlocal signal get suppressed at FMR condition, its magnitude is defined as $F_{2\pm}$ and $S_{2\pm}$ for the first and second harmonic nonlocal resistances (see green areas in Supplementary Fig.$\,$S3a and S3b). Also, when the nonlocal signal get enhanced at a slightly higher field than FMR condition, its magnitude is denoted as $F_{3\pm}$ and $S_{3\pm}$ (see red areas in Supplementary Fig.$\,$S3a and S3b). We define the amplitude of the suppression and enhancement as 
 	\begin{equation}
 	\Delta R^{1\omega}_{\textrm{asy}}=F_{\textrm{1}}-F_{0}
 	\end{equation}
 	
 	\begin{equation}
 	\Delta R^{1\omega}_{\textrm{sup}}=\frac{1}{2}(F_{\textrm{2}-}+F_{\textrm{2}+})-F_{0}
 	\end{equation}
 	
 	\begin{equation}
 	\Delta R^{2\omega}_{\textrm{sup}}=\frac{1}{2}(S_{\textrm{2}-}-S_{\textrm{2}+})-\frac{1}{2}(S_{0-}-S_{0+})
 	\end{equation}
 	
 	\begin{equation}
 	\Delta R^{1\omega}_{\textrm{enh}}=\frac{1}{2}(F_{\textrm{3}-}+F_{\textrm{3}+})-F_{0}
 	\end{equation}
 	
 	\begin{equation}
 	\Delta R^{2\omega}_{\textrm{enh}}=\frac{1}{2}(S_{\textrm{3}-}-S_{\textrm{3}+})-\frac{1}{2}(S_{0-}-S_{0+})
 	\end{equation}
 	for the first and second harmonic nonlocal resistance. The value of all F$_{\textrm{n}}$ and S$_{\textrm{n}}$ (n$\neq$0) are extracted by looking at the two or three points of peaks or dips, whose average values are used as illustrated in Fig.$\,$S3c and S3d. 
 	
 	We define an error of the extracted baseline resistance as 
 	\begin{equation}
 	\textrm{Error}(F_{0}, S_{0-}, S_{0+})=\frac{\sigma(F_{0}, S_{0-}, S_{0+})}{\sqrt{N}},
 	\end{equation}
 	where $N$ is the number of points used to calculate the mean values. $\sigma$ is the standard deviation, which is defined as 
 	\begin{equation}
 	\sigma=\sqrt{\frac{\sum_{N}^{i=1}(x_i-\bar{x})^2}{N-1}},
 	\end{equation}
 	where $x_i$ is the i$^{\textrm{th}}$ point and $\bar{x}$ is the mean value of all the points. The error of $F_{\textrm{n}}$ and $S_{\textrm{n}}$ (n$\neq$0) are defined as
 	\begin{equation}
 	\textrm{Error}(F_{1},F_{2\pm},F_{3\pm})=\frac{\sigma(F_{0})}{\sqrt{N^{\prime}}}
 	\end{equation}
 	\begin{equation}
 	\textrm{Error}(S_{2-},S_{3-})=\frac{\sigma(S_{0-})}{\sqrt{N^{\prime}}}
 	\end{equation}
 	\begin{equation}
 	\textrm{Error}(S_{2+},S_{3+})=\frac{\sigma(S_{0+})}{\sqrt{N^{\prime}}}
 	\end{equation}
 	where $\sigma(F_{0})$, $\sigma(S_{0-})$ and $\sigma(S_{0+})$ are the standard deviation from calculating the mean values of $F_{0}$, $S_{0-}$ and $S_{0+}$, respectively. $N^{\prime}$ is the number of points used to extract the value of F$_{\textrm{n}}$ and S$_{\textrm{n}}$ (n$\neq$0).

 	\hfill \break}

\end{document}


\renewcommand{\thesection}{S.\arabic{section}}
\renewcommand{\thesubsection}{\thesection.\arabic{subsection}}

\makeatletter 
\makeatother
\begin{center}
	\Large \uppercase{\textbf{Supplemental Material}}
\end{center}
\begin{center}
	\Large \textbf{Microwave control of thermal magnon spin transport}
\end{center}
\begin{center}
\large J. Liu*, F. Feringa, B. Flebus, L.J. Cornelissen, J.C. Leutenantsmeyer, R.A. Duine, B.J. van Wees
\end{center}

\begin{figure}[h!]
	\centering
	\includegraphics[width=0.55\linewidth]{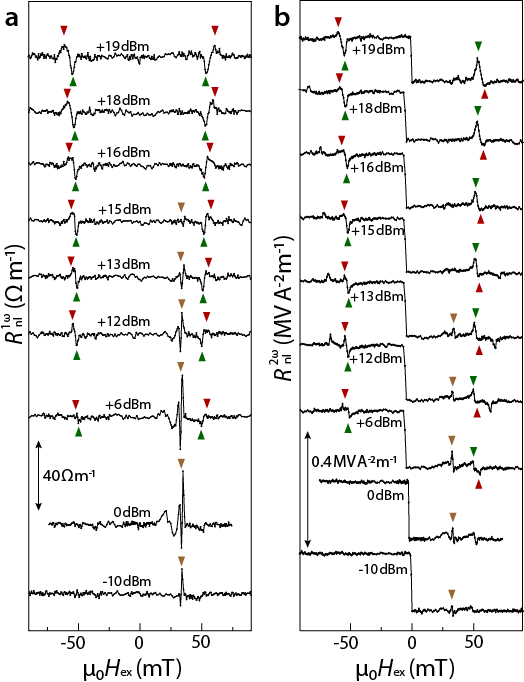}
	\caption{\textbf{Comparison between $R^{1\omega}_{\textrm{nl}}$ and $R^{2\omega}_{\textrm{nl}}$ at various rf power.} \textbf{a} First and \textbf{b} second harmonic nonlocal resistance as a function of external static field at different rf powers, from $-10\,$dBm to $+19\,$dBm with an rf frequency of 3$\,$GHz. Green and red triangles indicate the suppression and enhancement of signals at the FMR condition and at a higher field, respectively. The brown triangles point out the features only showing up at a positive field around 20$\,$mT smaller than the FMR condition. These asymmetric features only appear at relatively low $\textit{P}_{\textrm{rf}}$. The excitation current has a rms-amplitude of 200$\,\mu$A. The measured device has a injector-to-detector distance of 1$\,\mu$m on a YIG film of 210$\,$nm. The scale bars are \textbf{a} 40$\,\Omega\,$m$^{-1}$ and \textbf{b} 0.4$\,$MV$\,$A$^{-2}$m$^{-1}$.}
	\label{Prf_dependent_R}
\end{figure}

\begin{figure}[h!]
	\centering
	\includegraphics[width=0.99\linewidth]{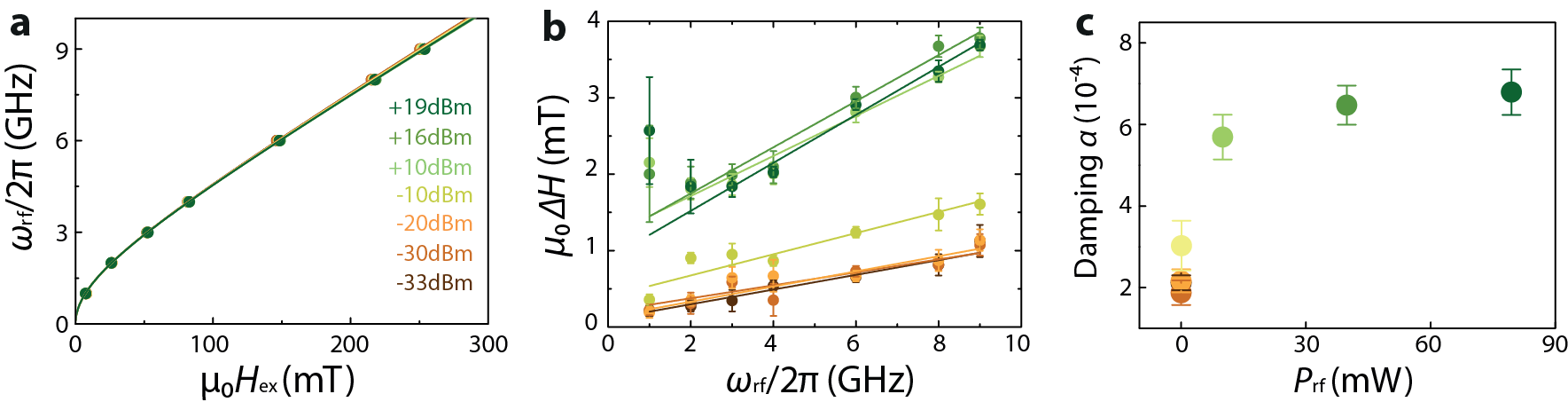}
	\caption{\textbf{Analysis of $S_{11}$ measurement.} \textbf{a} Rf resonance frequencies as a function of the static field. Different applied rf powers are indicated by the color from low (brown) to high (green) powers in \textbf{a}-\textbf{c}. The solid lines in \textbf{a} with corresponding colors are fits based on Kittel equation. \textbf{b} Linewidth of the absorption peak $\Delta H$ as a function of rf resonance frequencies. The solid lines with corresponding colors are the linear fits, from which we obtain the Gibert damping parameter $\alpha$, which is plotted against rf power in \textbf{c}.}
	\label{S11_SP_damping}
\end{figure}

\begin{figure}[h!]
	\centering
	\includegraphics[width=0.7\linewidth]{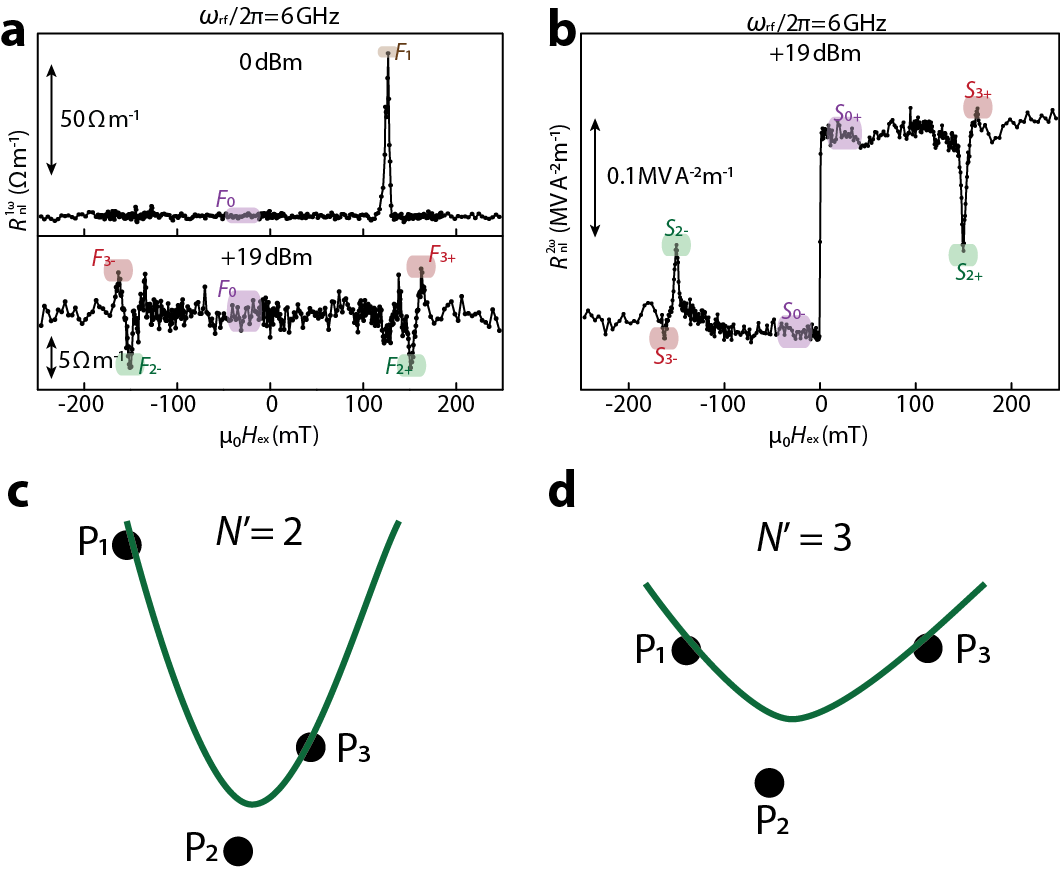}
	\caption{\textbf{Extracting the amplitude of $\Delta R_{\textrm{enh}}^{1\omega}$, $\Delta R_{\textrm{enh}}^{2\omega}$, $\Delta R_{\textrm{sup}}^{1\omega}$, $\Delta R_{\textrm{sup}}^{2\omega}$ and $\Delta R_{\textrm{asy}}^{1\omega}$.} Typical field-dependent \textbf{a} first and \textbf{b} second harmonic nonlocal resistances. Highlighted regions with different colors are used to extract the peak amplitudes and baseline resistances. The scale bars are 50$\,\Omega\,$m$^{-1}$ and 5$\,\Omega\,$m$^{-1}$ in the upper and lower panel of \textbf{a} for rf power at 0 and $+19\,$dBm, while in \textbf{b} the scale bar is 0.1$\,$MV$\,$A$^{-2}$m$^{-1}$. The rf field oscillates at 6$\,$GHz. \textbf{c}, \textbf{d} Visualization of the two situations of extracting the value of the peak amplitude. We take the mean value of the point P$_{2}$ and P$_{3}$ in \textbf{c}, where P$_{3}$ is much closer to the extremum (P$_{2}$) than P$_{1}$, while in \textbf{d} we take the mean value of points P$_{1}$, P$_{2}$ and P$_{3}$ since P$_{1}$ and P$_{3}$ are comparably close to P$_{2}$. The green lines are the lineshape of the peak feature based on extrema we calculated.}
	\label{peak_1}
\end{figure}

\begin{figure}[h!]
	\centering
	\includegraphics[width=0.99\linewidth]{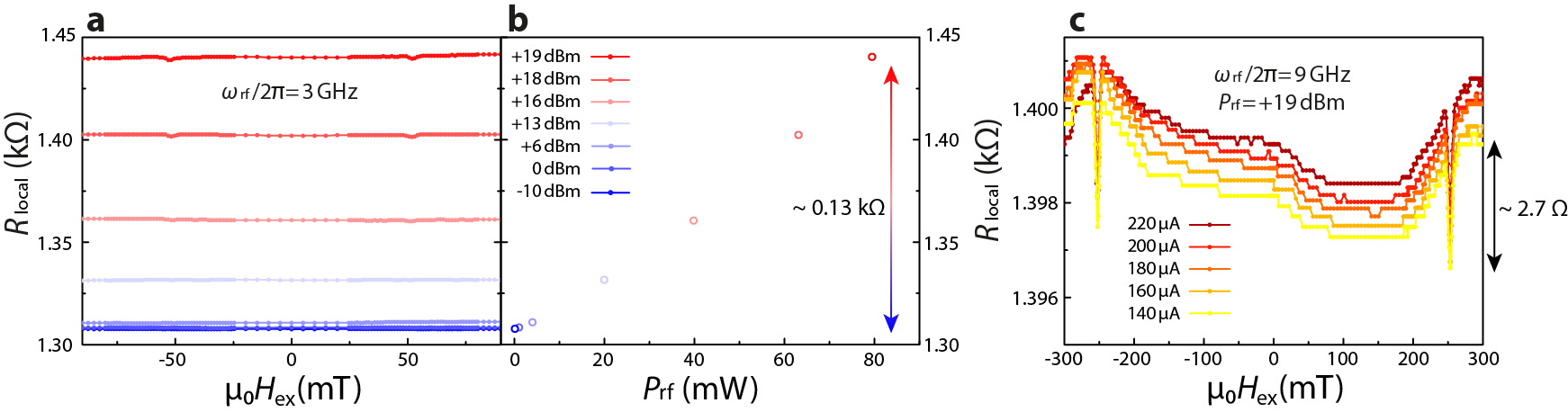}
	\caption{\textbf{Local measurement.} \textbf{a} Field dependent Pt resistance under the presence of an rf field oscillating at 3$\,$GHz at various rf powers ranging from -10$\,$dBm to 19$\,$dBm. \textbf{b} Rf-power dependent Pt base resistances, which are the mean of the resistances measured from -0.3$\mu_{0}H_{\textrm{FMR}}$ to 0.3$\mu_{0}H_{\textrm{FMR}}$. $H_{\textrm{FMR}}$ is the resonance field of $\sim50\,$mT for a 3$\,$GHz-oscillating rf field, at which we see a dip of the local resistance. The error bars lie within the data points. \textbf{c} Pt resistance measured at different excitation current of 140$\,\mu$A, 160$\,\mu$A, 180$\,\mu$A, 200$\,\mu$A and 220$\,\mu$A under the presence of an rf power of $\sim79.4\,$mW oscillating at 9$\,$GHz. The dips at $\sim250\,$mT correspond to the FMR condition for a 9$\,$GHz-oscillating rf field. The YIG thickness is 210$\,$nm. }
	\label{Pt_resistance}
\end{figure}

\begin{figure}[h!]
	\centering
	\includegraphics[width=0.99\linewidth]{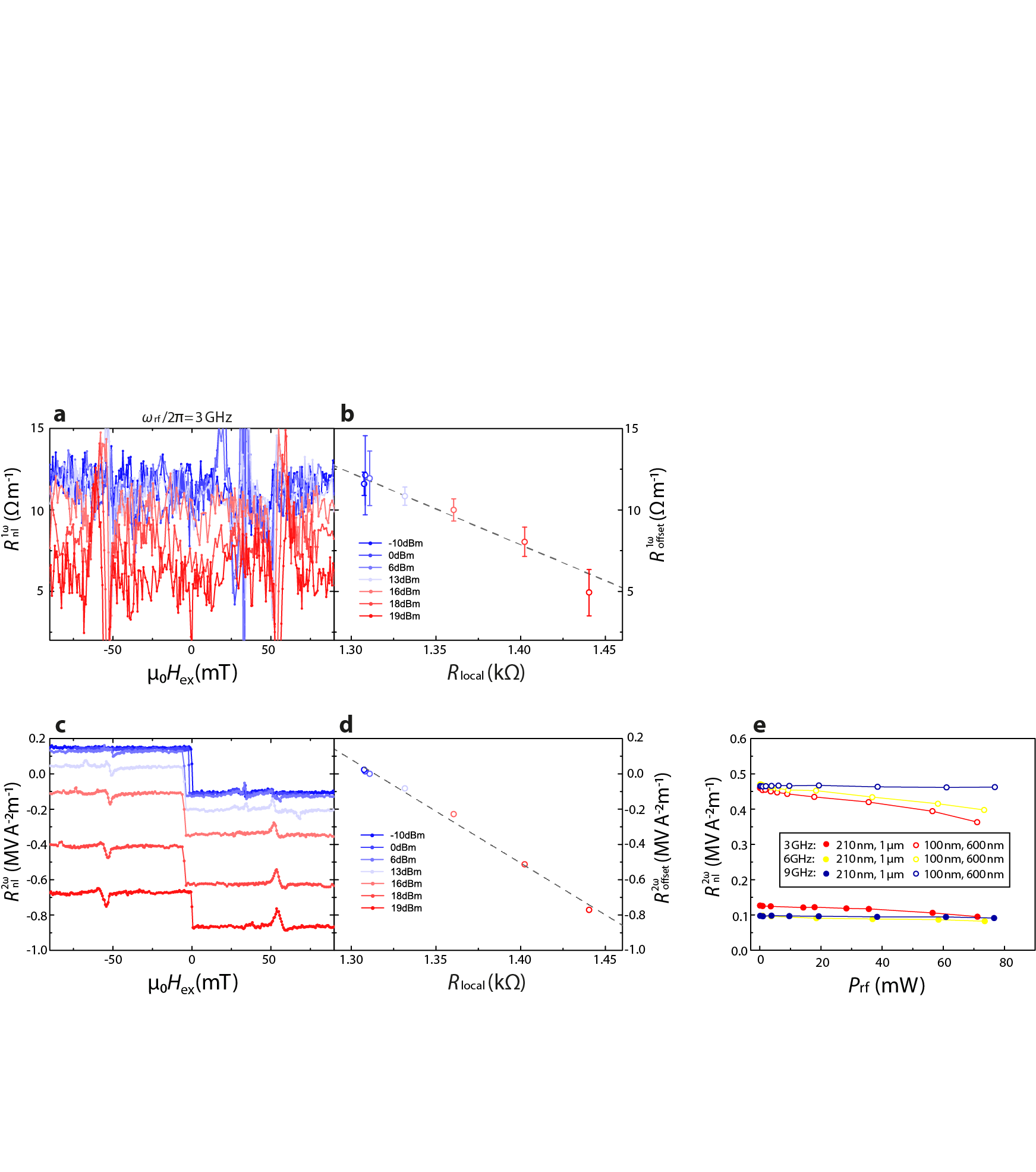}
	\caption{\textbf{Offset shifting of the nonlocal resistance.} Field dependent \textbf{a} first and \textbf{c} second harmonic signals under different rf powers from -10$\,$dBm to 19$\,$dBm (from blue to red). Same color code is used for applied rf power in \textbf{a}-\textbf{d}. In \textbf{a}, it is a zoom-in of the data near the baseline resistance. The offset of \textbf{b} first and \textbf{d} second harmonic signals is plotted against the measured local resistance at different rf powers with corresponding colors. The gray lines are linear fits. \textbf{e} $P_{\textrm{rf}}$-dependent magnitude of the second harmonic signals, i.e. half of the difference between base resistance of $R_{\textrm{nl}}^{2\omega}$ for the positive and negative fields. The $R_{\textrm{offset}}^{1\omega}$ is the mean values taken from 10$\,$mT to 40$\,$mT excluding the region around 33$\,$mT where there are the features due to the surface modes.}
	\label{R2w_baselineshift}
\end{figure}

\begin{figure}[h!]	
	\centering
	\includegraphics[width=0.9\linewidth]{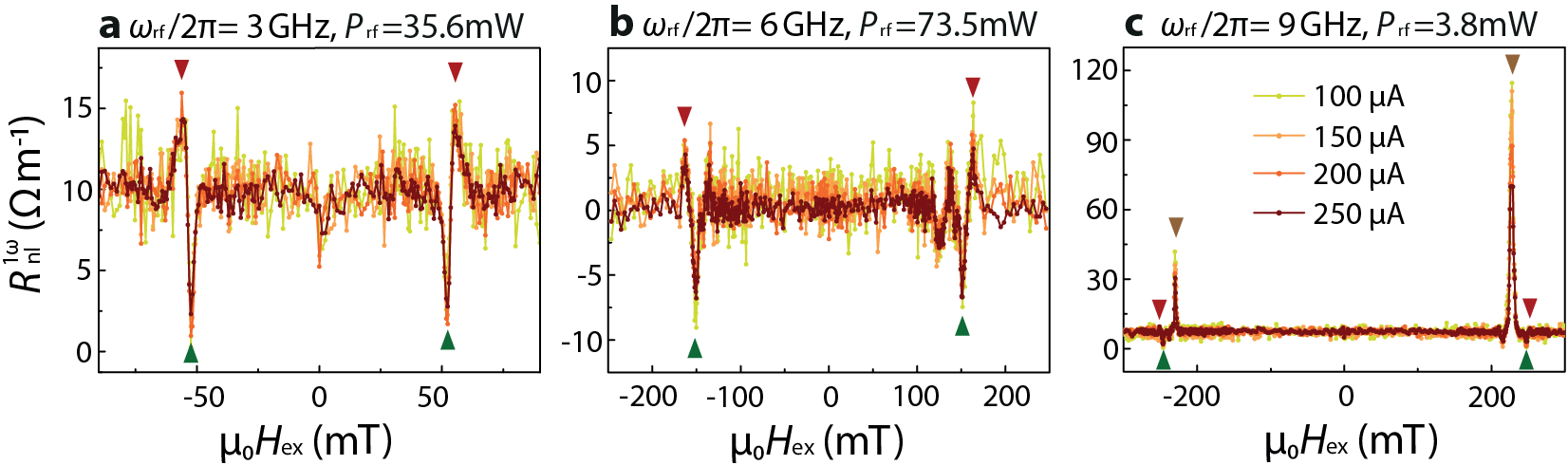}
	\caption{\textbf{Linearity check.} Excitation current dependent first harmonic results measured at rf frequencies of \textbf{a} 3$\,$GHz, \textbf{b} 6$\,$GHz and \textbf{c} 9$\,$GHz with different rf powers. The excitation current has rms-amplitude of 100$\,\mu$A, 150$\,\mu$A, 200$\,\mu$A and 250$\,\mu$A. Same color code is used for excitation current in \textbf{a}-\textbf{c}. The measured device is on top of 210-nm thick YIG film with injector-to-detector distance of 1$\,\mu$m. Note that in \textbf{c} due to the large response of the asymmetric features at positive field, the signals are larger than the signal limitation for 200$\,\mu$A and 250$\,\mu$A. }
	\label{Linearity}
\end{figure}

\begin{figure}[h!]
	\centering
	\includegraphics[width=0.9\linewidth]{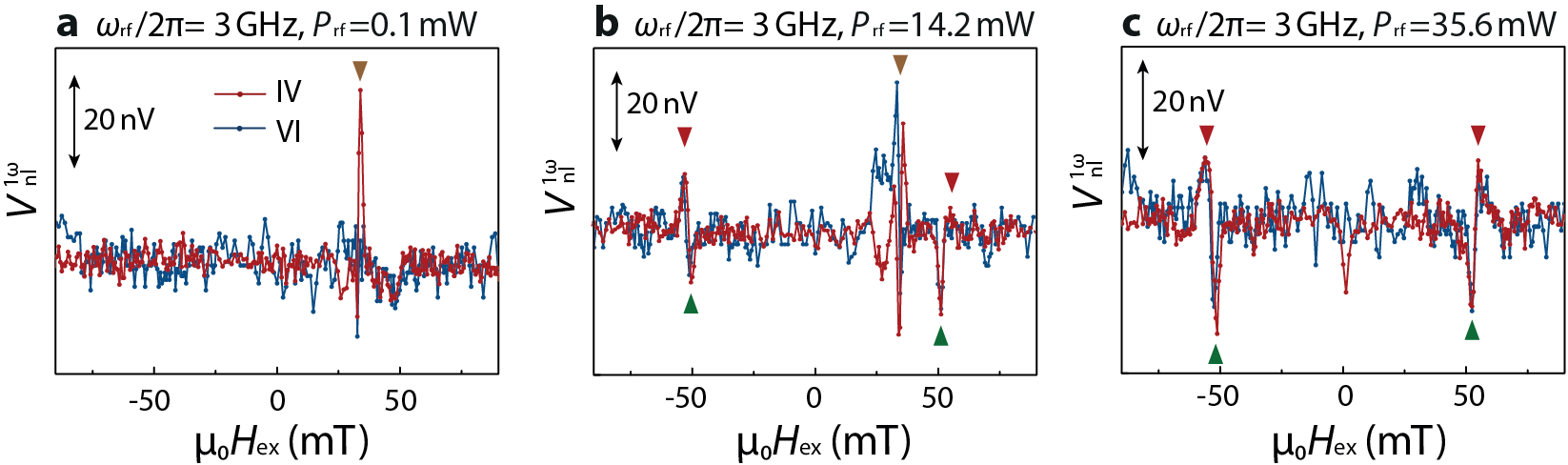}
	\caption{\textbf{Reciprocity check.} Comparison between the first harmonic results measured in IV (red) and VI (blue) configurations at rf powers of \textbf{a} 0.1$\,$mW, \textbf{b} 14.2$\,$mW and \textbf{c} 35.6$\,$mW. Same color code is used for measurement configuration in \textbf{a}-\textbf{c}. The measured device is on top of 210$\,$nm thick YIG film with injector-to-detector distance of 1$\,\mu$m. The applied rf field oscillates at 3$\,$GHz. The scale bars are 20$\,$nV.}
	\label{Reciprocity_rfpower}
\end{figure}

\begin{figure}[h!]
	\centering
	\includegraphics[width=0.55\linewidth]{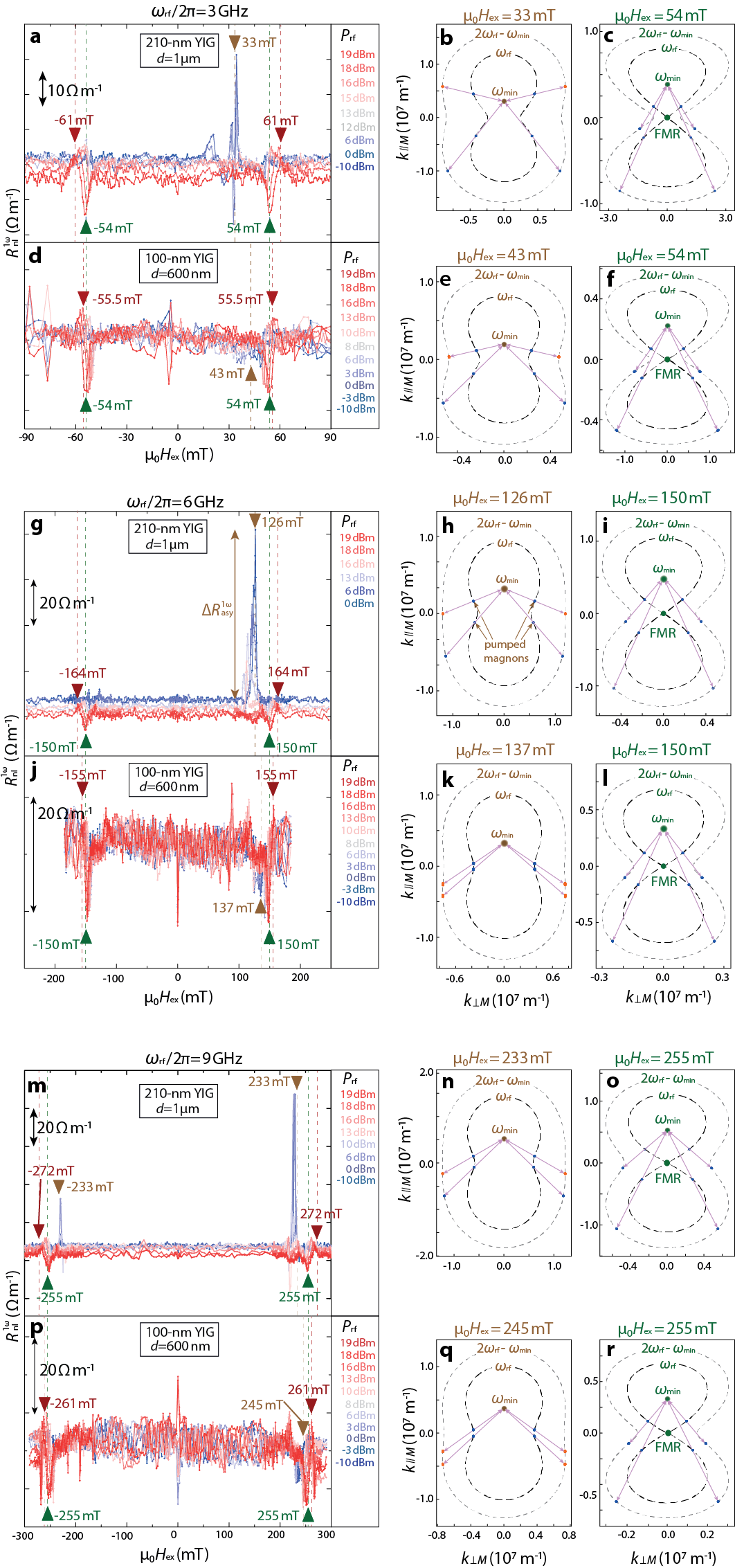}
	\caption{\textbf{Comparison between 210$\,$nm and 100$\,$nm thick YIG results at various rf frequencies and rf powers.} Typical field-dependent first harmonic nonlocal resistance of \textbf{a}, \textbf{g}, \textbf{m} 210$\,$nm and \textbf{d}, \textbf{j}, \textbf{p} 100-nm thick YIG at rf frequencies of \textbf{a}, \textbf{d} 3$\,$GHz, \textbf{g}, \textbf{j} 6$\,$GHz and \textbf{m}, \textbf{p} 9$\,$GHz. Applied $\textit{P}_{\textrm{rf}}$ ranges from -10 to 19$\,$dBm corresponding to the color spectrum from blue to red. The brown, green and red triangles with corresponding vertical dashed lines indicate special field positions. Relevant iso-frequency curves of \textbf{b}-\textbf{c}, \textbf{h}-\textbf{i},  \textbf{n}-\textbf{o}  210$\,$nm and \textbf{e}-\textbf{f}, \textbf{k}-\textbf{l}, \textbf{q}-\textbf{r}  100$\,$nm thick YIG dispersion relation: with magnon frequencies of $\omega_{\textrm{rf}}$ and $2\omega_{\textrm{rf}}-\omega_{\textrm{min}}$ in black and gray dashed lines, respectively. $k_{\perp M}$ and $k_{\parallel M}$ are wavevectors perpendicular and parallel to the in-plane magnetization, corresponding to the magnetostatic surface mode and backward volume mode. Four-magnon scattering process: two magnons with frequencies of $\omega_{\textrm{rf}}$ (blue dots) scatter with each other, giving rise to one magnon at band minima $\omega_{\textrm{min}}$ and one magnon with higher frequency of $2\omega_{\textrm{rf}}-\omega_{\textrm{min}}$. The purple double arrow lines represent the momentum conservation for the 4-magnon scattering process. Parameters used for plotting the dispersion relation are obtained from the Kittel fit of rf power reflection measurement, i.e. $S_{11}$ parameter: gyromagnetic ratio $\gamma=27.3\,$GHz/T and saturation magnetization $M_{\textrm{s}}=135\,$kA/m. We used exchange stiffness of $1\times10^{-39}\,$J$\,$m$^{2}$ \cite{kalinikos1986theory}.}
	\label{dispersion_for_three_fields_two_thicknesses}
\end{figure}

\pagebreak

\bibliography{reference}